\documentclass[aps,pre,reprint,floatfix,superscriptaddress,longbibliography]{revtex4-2}
\usepackage{cmap}
\usepackage{graphicx}
\usepackage{hyperref}
\usepackage{amsmath,amssymb}

\begin{document}

\title{Electrical conductivity of crack-template-based transparent conducting films: mean-field approximation, effective-medium theory, and simulation}

\author{Yuri Yu. Tarasevich}
\email[Corresponding author: ]{tarasevich@asu-edu.ru}
\affiliation{Laboratory of Mathematical Modeling, Astrakhan Tatishchev State University, Astrakhan, 414056, Russia}

\author{Andrei V. Eserkepov}
\email{dantealigjery49@gmail.com}
\affiliation{Laboratory of Mathematical Modeling, Astrakhan Tatishchev State University, Astrakhan, 414056, Russia}

\author{Irina V. Vodolazskaya}
\email{vodolazskaya\_agu@mail.ru}
\affiliation{Laboratory of Mathematical Modeling, Astrakhan Tatishchev State University, Astrakhan, 414056, Russia}

\date{\today}

\begin{abstract}
In this work, crack-template-based transparent conducting films were modeled as networks corresponding to the edges of a two-dimensional Poisson--Voronoi diagram. Two types of networks were considered: the original one, in which the conductance of each edge was inversely proportional to its length, and the effective one, in which all edges had the same conductance obtained from the effective-medium theory. The mean-field approximation was used for analytical evaluation of the electrical conductivity. Direct numerical calculations for the Poisson--Voronoi diagram showed that the mean-field approximation overestimated the effective conductivity of the original network by approximately 13\%, and of the effective network by 79\%. In addition, a honeycomb network with an edge conductance distribution corresponding to the Poisson--Voronoi diagram was studied: for it, the predictions of the effective-medium theory turned out to be more accurate than for the Poisson--Voronoi diagram, which was explained by the greater structural homogeneity of the periodic honeycomb lattice. The results indicate that, when modeling crack-template-based transparent conducting films, the application of the mean-field approximation may lead to significant errors if the resistance of individual conductors is not simply proportional to their length. This possibility is discussed as a motivation for future studies of hierarchical cracks with variable width, which are not directly investigated here.
\end{abstract}

\maketitle

\section{Introduction\label{sec:intro}}
Crack-template-based (CTB) transparent conducting films (TCFs) are widely used in various devices: transparent heaters, shielding devices, electroluminescent devices, solar cells, smart windows, etc.~\cite{Cama2025a,Cama2025}. While in the case of TCFs based on metal nanowires or carbon nanotubes, contacts make a significant contribution to the resistance, CTB TCFs belong to the so-called seamless systems, i.e., they completely lack contact resistance. In some cases, for example, when used as transparent heaters on automotive windows, the appearance of a diffraction pattern is unacceptable; therefore, the use of periodic templates is ruled out. Because crack templates have no regular structure, CTB TCFs do not create a pronounced diffraction pattern and provide uniform illumination. Moreover, a key requirement for transparent heaters is uniformity of heating and a short response time.

A common approach to modeling crack templates is to use Poisson--Voronoi diagrams on a plane~\cite{Zeng2020,Kim2022,Tarasevich2023APL,Esteki2023,Qiang2024,Tarasevich2025}. A theoretical study of the influence of geometry on the electrothermal and optical properties of a metallic mesh structure of CTB TCFs using a coupled electrothermal model was carried out in~\cite{Zeng2020}. In~\cite{Kim2022}, a geometric modeling approach for networks formed by crack templates was proposed, and simulations were used to compute their wavelength- and angle-dependent optical transmittance and sheet resistance. In~\cite{Esteki2023}, an in-depth computational study of the thermo-electro-optical properties of seamless nanowire networks was conducted to understand their geometric features using proprietary computational implementations and a coupled electrothermal model built in COMSOL Multiphysics. In~\cite{Qiang2024}, a coupled electrothermal model and an electromagnetic model were developed to investigate the properties of an embedded metal mesh in a polymer matrix.

In~\cite{Tarasevich2023APL}, within the mean-field approximation (MFA), a formula was obtained for the electrical conductance of a network corresponding to the edges of a two-dimensional Poisson--Voronoi diagram (2D PVD)
\begin{equation}\label{eq:SigmaAPL}
G^\text{PVT}_\text{MFA} = \sigma_0 A\sqrt{\frac{n}{3}},
\end{equation}
where $\sigma_0$ is the electrical conductivity of the material, $A$ is the cross-sectional area of the conductor, and $n$ is the number of edges per unit area.

Using the MFA, the effective electrical conductivity of dense two-dimensional (2D) random resistor networks (RRNs) created using a Poisson--Voronoi tessellation was studied~\cite{Tarasevich2023APL}. These networks can mimic, for example, CTB TCFs. Although such random resistor networks are isotropic and on average homogeneous, local fluctuations of the number of edges per unit area are inevitable. An analytical dependence of the effective conductivity on the number of conducting edges per unit area was proposed. This formula was compared with other formulas proposed in the literature, as well as with direct calculations of the effective electrical conductivity of dense random resistor networks. The comparison showed that all theoretical estimates overestimate the effective conductivity even for very dense random resistor networks, for which the MFA is expected to be sufficiently accurate.


Although both the MFA and the EMT have been applied to random resistor networks mimicking CTB TCFs, the accuracy of these approximations and the reasons for their deviations from direct numerical calculations remain poorly understood. In particular, it is unclear whether the discrepancies arise from the approximations themselves, from the neglect of potential fluctuations at the network nodes required by Kirchhoff's current law, or from the structural properties of the Poisson--Voronoi diagram. Moreover, no systematic comparison of the MFA, EMT, and direct numerical calculations has been performed on the same model system. In this work, we address this gap by (i) deriving MFA expressions that explicitly include potential fluctuations, (ii) applying EMT to the same network, and (iii) comparing both approximations with direct numerical calculations for the original and effective networks. This comparison allows us to identify the limits of applicability of each method and to explain the origin of their errors. Understanding these limits is important because it provides guidance for choosing the appropriate method when modeling real CTB TCFs, helps to identify which structural and physical features are responsible for the deviations, and may stimulate the development of more accurate analytical approaches.

The remainder of this paper is organized as follows. Section~\ref{sec:methods} describes the models and simulation methods used. Section~\ref{sec:results} presents analytical estimates and computer simulation results. Section~\ref{sec:concl} summarizes the main findings.

\section{Methods\label{sec:methods}}
\subsection{General remarks}

Image processing results of published photographs of CTB TCFs showed that the typical vertex degree in crack patterns is 3~\cite{Tarasevich2023PRE}. Dead ends correspond to a small number of nodes with degree 1; the edges incident to them do not contribute to the electrical conductivity of the network. Moreover, the boundaries of the photographs create fictitious nodes with degree 1; these fictitious nodes do not correspond to any real node. A small number of nodes with degree 2 essentially corresponds to bends in the cracks. A small number of nodes with degree greater than 3 should be regarded as an image processing artifact due to the modest resolution of the photographs, when two or more nodes are mistaken for a single node, since simple mechanical arguments suggest that X-shaped cracks are extremely unlikely. Thus, networks with degree 3 (3-regular networks) can be used to mimic CTB TCFs.

Since TCFs are used in low-voltage electronics to prevent fogging and icing of windows, the typical temperature rise amounts to several tens of degrees. To a first approximation, the electrical conductivity of the metal can be considered temperature-independent. Our focus is on CTB TCFs based on so-called seamless nanowire networks. Such networks can be obtained, for example, by filling crack templates based on microscopic crack networks with metal.

Note that, following the literature on random resistor networks, we refer to the conductance of the entire network as its effective conductivity, since the network is treated as an effective medium. To avoid confusion, the conductance of an individual edge is always referred to as conductance.

\subsection{Poisson--Voronoi diagram\label{subsec:VT}}

A Voronoi diagram is a tessellation of space into regions close in Euclidean distance to each of a given set of objects (see, e.g.,~\cite{Okabe2000}). In our study, we deal with random Voronoi diagrams (Poisson--Voronoi diagrams) on a plane (2D PVD). In this particular case, points (seeds) are randomly distributed within a bounded region on the plane. In a 2D PVD, the degree of each vertex is 3, and the average number of edges per cell is 6~\cite{Meijering1953}. Thus, from a graph theory perspective, the edges of a 2D PVD form a 3-regular planar graph. Since the crack patterns exhibit similar properties, the 2D PVD appears to be a reasonable and useful mathematical model of the crack pattern~\cite{Zeng2020,Kim2022,Tarasevich2023PRE,Esteki2023,Tarasevich2023APL}.

Therefore, in our study we use a network whose edges correspond to the edges of a 2D PVD. For a 2D PVD, the probability density function (PDF) of edge lengths is known in quadratures~\cite{Muche1996,Brakke2005}; the PDF for unit intensity of the Poisson process (seed concentration) was calculated in~\cite{Brakke2005}. For an arbitrary intensity $n_\text{s}$ of the Poisson process, the PDF of edge lengths of a 2D PVD can be obtained by scaling from the PDF for unit intensity
\begin{equation}\label{eq:PDFscaling}
f_L\left(l;n_\text{s}\right) = \sqrt{n_\text{s}} f_L(\sqrt{n_\text{s}} l;1).
\end{equation}
Indeed, if the dimensions are scaled by a factor of $\sqrt{n_\text{s}}$, the area of the region under consideration increases by a factor of $n_\text{s}$, and the seed concentration (intensity of the Poisson process) changes by a factor of $n_\text{s}^{-1}$.

Several important relations are known for a 2D PVD. Since, according to~\cite{Meijering1953}, for a unit-intensity Poisson process in such diagrams the average number of edges per cell is 6 and the average perimeter length is 4 in units of the mean seed spacing, the mean edge length (first moment) is
\begin{equation}\label{eq:EL}
\mathbb{E}[L] = \frac{2}{3}.
\end{equation}
For the second moment, only a result obtained by numerical integration of an analytical formula is known~\cite{Brakke2005},
\begin{equation}\label{eq:EL2}
  \mathbb{E}[L^2] = C,\quad \text{where } C \approx 0.63007.
\end{equation}
For an arbitrary intensity $n_\text{s}$ of the Poisson process, the scaling relations \(\mathbb{E}[L] \propto n_\text{s}^{-1/2}\) and \(\mathbb{E}[L^2] \propto n_\text{s}^{-1}\) are direct consequences of the scaling property of the Poisson process.

\begin{figure}[!htb]
 \centering
 \includegraphics[width=\columnwidth]{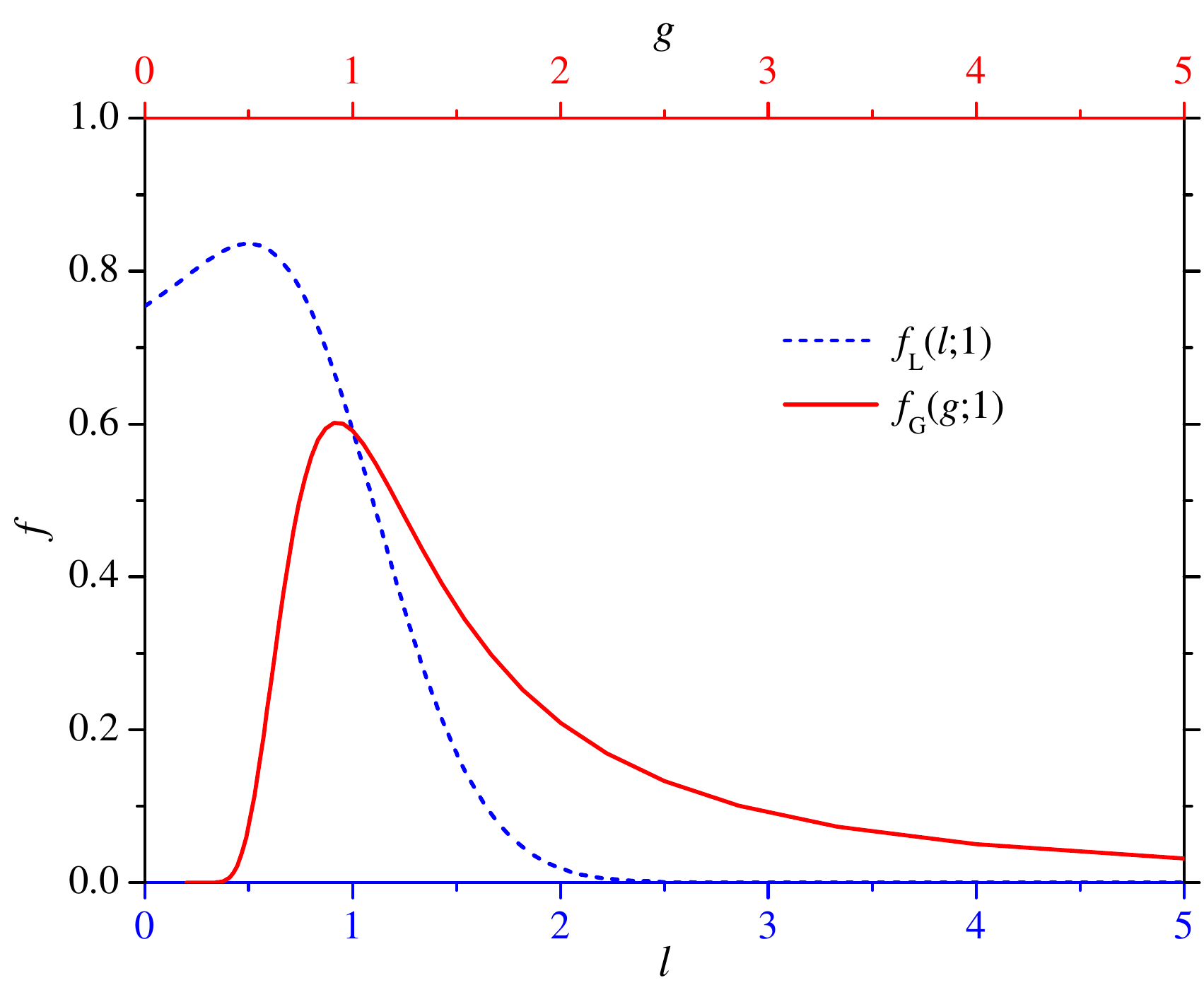}
 \caption{PDF of the length $l$ and electrical conductance $g$ of edges in a network corresponding to the edges of a 2D PVD~\eqref{eq:appfG} for unit intensity of the Poisson process (one seed per unit area). The PDF of edge lengths was taken from~\cite{Brakke2005}, whereas the PDF of electrical conductance was obtained using~\eqref{eq:appfG}.}
 \label{fig:fvsg}
\end{figure}

When the edge length $l$ has a PDF $f_L(l)$, the PDF of the electrical conductance $f_G(g)$ can be found by noting that $g = \lambda/l$, where $\lambda = \sigma_0 A$ is the conductance per unit length. Then
\begin{equation}\label{eq:appfG}
 f_G( g;n_\text{s} ) = \frac{ \lambda }{ g ^2} f_L\left( \frac{ \lambda }{ g };n_\text{s} \right).
\end{equation}
Figure~\ref{fig:fvsg} shows the PDF of edge lengths for a 2D PVD at unit intensity of the Poisson process, as well as the corresponding PDF of edge conductances~\eqref{eq:appfG}.

\subsection{Model of the conducting network}

Consider a $z$-regular ($z>2$) planar graph embedded in the plane, consisting of a single connected component. This graph defines the topology of both networks considered below; they differ only in the assignment of edge conductances.

Consider two conducting networks constructed on the basis of this graph.
\begin{enumerate}
\item Assume that the edges of the graph under consideration are conductors with constant and identical cross-section $A$, and the electrical conductivity of the conductor material is $\sigma_0$. We will call such a network the \emph{original network}. This is the physical model of a CTB TCF: it corresponds to a crack template filled with a conductor of uniform cross-section, so that the conductance of an edge is inversely proportional to its length.
\item Assume that the edges of the graph under consideration are conductors, and all edges have the same electrical conductance equal to $g_\text{m}$. We will call such a network the \emph{effective network}. This network is not a separate physical object but a methodological construct built on the same graph as the original network. It serves (i) to isolate the effect of conductance disorder from the effect of network topology and (ii) to provide the uniform network for which EMT directly predicts the edge conductance: EMT replaces the distribution of edge conductances of the original network by a single effective value $g_\text{m}$, and the effective network is precisely the result of this replacement.
\end{enumerate}
Remarks concerning the effective network: (i) at this stage we deliberately do not specify the value of $g_\text{m}$ or the method for finding it, since for the main part of the theoretical consideration it is only essential that it be a constant; (ii) regardless of real CTB TCFs, the effective network can be obtained from the original network, if necessary, by replacing conductors of identical cross-section $A$ with conductors whose cross-section is related to the conductor length by the following relation
\begin{equation}\label{eq:AMFA}
A^\text{eff} = \frac{g_\text{m}}{\sigma_0}  l.
\end{equation}

Note that in theoretical analysis it is customary to use the number of conductors per unit area $n$, whereas in practice the areal mass density (amd) of metal per unit area of the conducting film is usually measured. The relationship between these two quantities can easily be found. Indeed, the mass of a single conductor of length $l$ and cross-section $A$ is $\rho A l$ (where $\rho$ is the density of the conductor material), so the total mass of metal per unit area is
\begin{equation}\label{eq:amd}
\mu = \int\limits_0^\infty \rho A  l  \, n_L( l )\,\mathrm{d} l  = \rho A n \langle  l  \rangle.
\end{equation}

Since random crack networks are often modeled using a 2D PVD, we will separately consider the case where the 3-regular network corresponds to the edges of a 2D PVD. Note that although a 2D PVD is a 3-regular graph whose faces have on average 6 nodes, it is not a perfect honeycomb network.

To assess the effect of network imperfection, we will additionally consider the electrical conductivity of honeycomb networks in which the edge resistances are distributed according to the PDF of edge lengths in a 2D PVD~\cite{Brakke2005}. The electrical conductivity of a periodic honeycomb network with identical edge conductances $g_\text{m}$ can easily be obtained analytically:
\begin{equation}\label{eq:Ghex}
 G^\text{hex}= \frac{g_\text{m}}{\sqrt{3}}.
\end{equation}

\subsection{Details of simulation}

The simulation parameters were defined as follows. In the case of Poisson--Voronoi diagram, the computational domain was a square region $L \times L$ with side length \(L = 32\). Number of seeds was varied from 1000 to 5000 in increments of 200, that is, the intensity of the Poisson process (seed concentration) was approximately varied from 1 to 5. Because standard Voronoi tessellation algorithms generate the diagram on the entire plane by introducing a point at infinity rather than restricting it to a prescribed domain, the resulting tessellation was truncated to the domain boundaries. All edges and vertices were examined. An edge was discarded if both of its vertices lay outside the domain boundaries, including cases in which the edge passed through a corner of the domain. If one vertex lay outside the boundaries while the other lay inside the domain, the intersection point with the boundary was determined by successively testing each boundary. Once the intersection point was found, the coordinates of the exterior vertex were replaced by those of the intersection point with the boundary. This procedure yielded an accurate truncation of the tessellation to the computational domain.

\begin{figure}[htb]
  \centering
  \includegraphics[width=\columnwidth]{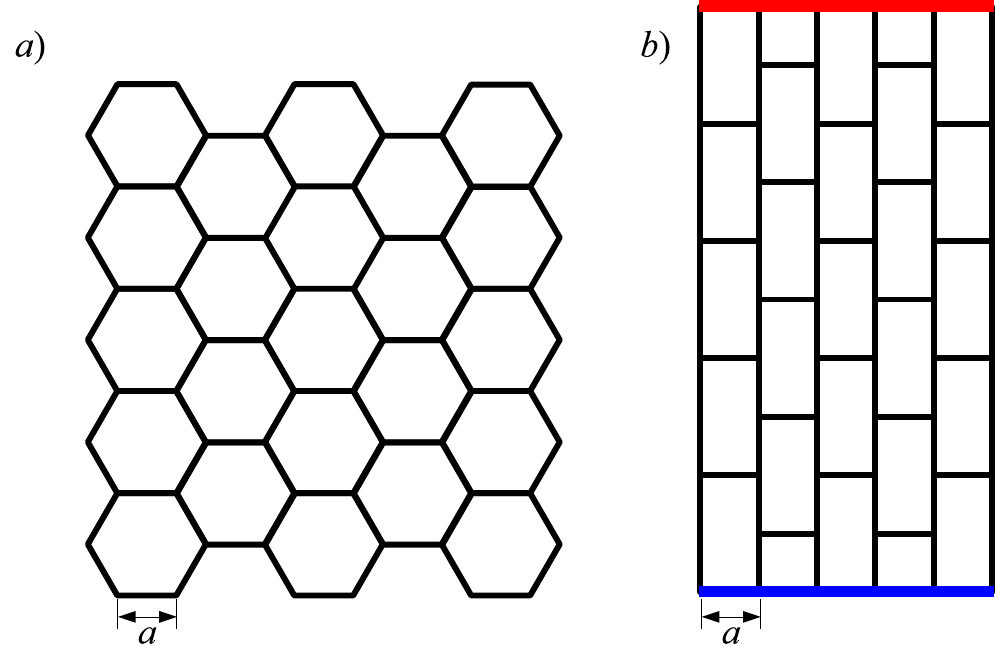}
  \caption{a) Honeycomb lattice; b) ``brickwork'' lattice, topologically equivalent to the honeycomb lattice.}\label{fig:network6}
\end{figure}

In the case of a hexagonal lattice, it is technically simpler to perform calculations on a topologically equivalent ``brickwall'' lattice (see Fig.~\ref{fig:network6}). Since the ``brickwall'' lattice is rectangular ($L_x \times L_y$), the calculated conductances depend on direction. In such cases, one usually determines the sheet resistance, which, for a square sample, coincides with the resistance, and for a rectangular one is given by
$$
R_\Box = R_x \frac{L_y}{L_x} = R_y \frac{L_x}{L_y},
$$
where $R_x$ and $R_y$ are the resistances along the $x$-axis and the $y$-axis, respectively. $G = R^{-1}_\Box$ is the desired effective conductivity. Since, in the case of a finite-size system with random edge resistances, generally speaking $R_x \ne R_y$, we used averaging over the two directions
$$
\langle R_\Box \rangle = \frac{1}{2}\left( R_x \frac{L_y}{L_x} + R_y \frac{L_x}{L_y}\right).
$$

To calculate the electrical conductivity of the networks under consideration, we attached a pair of superconducting busbars to two opposite boundaries of the domain such that the voltage $U_0$ was applied either along the $x$-axis or along the $y$-axis. In the simulations, this implied that all vertices of the Voronoi diagram lying on the corresponding side of the domain were merged into a single vertex. By applying Ohm's law to each resistor (network edge) and Kirchhoff's current law to each node (network vertex), we obtained a system of linear equations, which was solved numerically. The total electric current flowing through the system, divided by the applied voltage, defines the conductance of the system. Since only square domains were considered, this quantity is independent of the system size \(L\) and can therefore be interpreted as the effective electrical conductivity.

For each seed concentration, the effective conductivity was averaged over 50 independent realizations and over both directions. Such a number of independent realizations is sufficient to ensure a standard error of the mean on the order of 0.1\% in the calculation of the electrical conductivity. Throughout this manuscript, the reported uncertainties are standard errors of the mean (SEM) over the 50 independent realizations, unless explicitly stated otherwise. In the plots, the SEM is on the order of the marker size.

\subsection{Effective-medium theory\label{subsec:EMT}}

Effective-medium theory~\cite{Bruggeman1935} is an approach often used to predict physical properties, e.g., the electrical conductivity of random systems, including RRNs~\cite{Kirkpatrick1973,Clerc1990,Luck1991,Sahimi1997,OCallaghan2016}. The main ideas of applying EMT to networks with a regular structure and varying edge conductances are presented in~\cite{Kirkpatrick1971,Kirkpatrick1973,Joy1978,Joy1979,Clerc1990}. Alternatively, a more formal and general treatment based on Foster's theorem is possible~\cite{Foster1961,Marchant1979}.
The idea is to find an effective edge conductance $g_\text{m}$ such that when all edges of the network are replaced by edges with identical conductance $g_\text{m}$, the conductance measured between any pair of nearest nodes is close to the average conductance measured between any pair of nearest nodes of the original network. For a $z$-regular network whose edge conductances follow a distribution $f_G(g)$, the effective conductance $g_\text{m}$ is found from the integral equation
\begin{equation}\label{eq:EMT}
\int f_G(g) \frac{g_\text{m} - g}{g + (z/2 - 1) g_\text{m}} \, \mathrm{d}g = 0.
\end{equation}
Here $f_G(g)$ is the edge-conductance distribution of the original network, and the solution $g_\text{m}$ defines the effective network introduced above. A limitation of EMT is that it generally does not yield the conductivity of the entire network. For some periodic lattices, e.g., square, honeycomb, and triangular, the relationship between the edge conductance and the conductivity of the entire lattice can easily be found~\cite{Kirkpatrick1971,Kirkpatrick1973,Joy1978,Joy1979}. However, for an arbitrary $z$-regular network, this relationship is far from obvious. Although EMT makes it possible to significantly simplify the problem by turning a random regular network into a uniform regular network, such a simplification is not always sufficient to find the network conductivity.

\section{Results\label{sec:results}}

\subsection{Theoretical treatment}

\subsubsection{Mean-field approximation}

Suppose a uniform electrostatic field with strength $E$ is applied, and the field vector lies in the plane of the conducting network. If the network is macroscopically homogeneous, one can expect that the electric potential varies along the network almost linearly.

The MFA is based on considering a single conductor placed in the field created by all the other conductors. In the simplest case, the potential of this field is assumed to depend linearly on the coordinate. However, the assumption of strict linearity of the potential contradicts Kirchhoff's current law.

Consider an arbitrary vertex of a $z$-regular network. Exactly $z$ edges meet at this vertex. Let the edge vectors emanating from the vertex be \(\mathbf{l}_1, \mathbf{l}_2, \dots, \mathbf{l}_z\) (directed from the vertex to its neighbors). The external field \(\mathbf{E}\) creates a linear potential, i.e., \(u_0(\mathbf{r}) = - (\mathbf{E} \mathbf{r})\).

If the potential were exactly linear, the current from the vertex along the $k$-th edge, whose endpoints are vertices $V$ and $W$, would be \(i_k^{(0)} = g_k (u_0(V) - u_0(W)) = g_k (\mathbf{E} \mathbf{l}_k)\). The sum of currents at the vertex (currents flowing out of the vertex are taken as positive) is
\begin{equation}\label{eq:Kirch1}
\sum_{k=1}^z i_k^{(0)} = \mathbf{E} (g_1\mathbf{l}_1 + g_2\mathbf{l}_2+\dots + g_z\mathbf{l}_z).
\end{equation}
In general, this sum is not zero. Even in the simplest case where all edge conductances are identical and can be factored out (the effective network), the vector sum of the edge vectors emanating from a vertex in a $z$-regular network does not vanish; consequently, Kirchhoff's current law is not satisfied.

In the case of the original network, the edge conductance is \(g_k = \lambda / l_k\), so under a linear potential the current is \(i_k^{(0)} = (\lambda / l_k) (\mathbf{E} \mathbf{l}_k) = \lambda E \cos\alpha_k\), which does not depend on length. Then the sum of currents at the vertex is
\begin{equation}\label{eq:Kirch-orig}
\sum_{k=1}^z i_k^{(0)} = \lambda E \sum_{k=1}^z \cos\alpha_k.
\end{equation}
Although the angles between the edges at any vertex sum to \(2\pi\), this does not mean that the sum of their cosines automatically equals zero; hence, Kirchhoff's current law is again not satisfied.

Thus, to satisfy Kirchhoff's current law at each vertex, microscopic potential fluctuations at the network vertices are required. For Kirchhoff's current law to hold, the potential of vertex \(V\) must deviate from the linear value: \(u_V = u_0(V) + \Delta u_V\).

\subsubsection{Thermal power of the original network}

The current along the $k$-th edge, whose endpoints are vertices $V$ and $W$, is
\begin{equation}\label{eq:ik-orig}
i_k = \frac{\lambda}{l_k} ((\mathbf{E} \mathbf{l}_k) + \delta u_k) = \lambda E \cos\alpha_k + \frac{\lambda}{l_k} \delta u_k.
\end{equation}
Here $\delta u_k = \Delta u_V - \Delta u_W$ is the difference of potential fluctuations at the ends of the $k$-th edge.
By Kirchhoff's current law,
\begin{equation}\label{eq:sumik-orig}
\sum_k i_k = \lambda E \sum_k \cos\alpha_k + \lambda \sum_k \frac{\delta u_k}{l_k} = 0,
\end{equation}
where the summation is over all edges meeting at the node under consideration. Consequently,
\begin{equation}\label{eq:zero-orig}
\sum_k \left(E \cos\alpha_k + \frac{\delta u_k}{l_k} \right) = 0.
\end{equation}
The power dissipated in the $k$-th edge is
\begin{multline}\label{eq:qkorig}
q_k = \frac{\lambda}{l_k} \left((\mathbf{E} \mathbf{l}_k) + \delta u_k\right)^2\\
= \lambda \left(E^2 l_k \cos^2\alpha_k + 2 E \delta u_k\cos\alpha_k + \frac{(\delta u_k)^2}{l_k}\right).
\end{multline}
Assuming that the edge orientations and lengths are independent random variables, we find the average thermal power dissipated in one edge of the network:
\begin{multline}\label{eq:meanqk-orig}
\langle q \rangle = \lambda E^2 \\ \times \left(\langle l \rangle \left\langle \cos^2\alpha \right\rangle + \frac{2}{E} \left\langle\delta u \cos\alpha\right\rangle + \frac{1}{E^2}\left\langle\frac{(\delta u)^2}{l_k}\right\rangle\right).
\end{multline}

Multiplying by the number of edges per unit area, we obtain the average thermal power dissipated per unit area:
\begin{multline}\label{eq:meanQ-orig}
\langle Q \rangle = \lambda n E^2 \\ \times \left( \langle l \rangle \left\langle\cos^2\alpha \right\rangle + \frac{2}{E} \left\langle\delta u \cos\alpha\right\rangle + \frac{1}{E^2}\left\langle\frac{(\delta u)^2}{l_k}\right\rangle \right).
\end{multline}
From Joule's law, the electrical conductivity of the network equals the factor multiplying $E^2$ in Eq.~\eqref{eq:meanQ-orig}:
\begin{multline}\label{eq:G-orig}
G^\text{orig}_\text{MFA} = \lambda n \\ \times \left( \langle l \rangle \left\langle\cos^2\alpha\right\rangle + \frac{2}{E} \left\langle\delta u \cos\alpha\right\rangle + \frac{1}{E^2}\left\langle\frac{(\delta u)^2}{l_k}\right\rangle \right).
\end{multline}
The second and third terms in parentheses are corrections to the ``straightforward'' MFA, which does not take fluctuations into account. In general, these corrections are nonzero.

\subsubsection{Thermal power of the effective network\label{subsubsection:HeatEffNetwork}}

Similarly, we find the thermal power dissipated in the effective network.
The current along the $k$-th edge, whose endpoints are vertices $V$ and $W$, is
\begin{equation}\label{eq:ik-eff}
i_k = g_\text{m} (u_V - u_W) = g_\text{m} \left((\mathbf{E} \mathbf{l}_k) + \delta u_k\right).
\end{equation}
Kirchhoff's current law gives
\begin{equation}\label{eq:Kirch-eff}
\sum_k \left( E l_k \cos \alpha_k + \delta u_k \right) = 0 .
\end{equation}
The power dissipated in this edge is
\begin{multline}\label{eq:qk-eff}
q_k = g_\text{m} (E l_k \cos\alpha_k + \delta u_k)^2\\ = g_\text{m} \left(E^2 l_k^2 \cos^2\alpha_k + 2\delta u_k E l_k \cos\alpha_k + (\delta u_k)^2\right).
\end{multline}
Assuming that edge orientations and lengths are independent, the average power dissipated in an arbitrarily chosen edge is
\begin{multline}\label{eq:meanq-eff}
\langle q \rangle = g_\text{m} \\ \times \left( \langle l^2 \rangle \langle \cos^2\alpha \rangle E^2 + 2 E \langle l \delta u \cos\alpha \rangle + \langle(\delta u)^2 \rangle \right).
\end{multline}
The total power dissipated per unit area of the system under consideration is
\begin{multline}\label{eq:Q-eff}
Q = g_\text{m} n \\ \times \left(\langle l^2 \rangle \langle \cos^2\alpha \rangle E^2 + 2 E \langle l \delta u \cos\alpha \rangle + \langle(\delta u)^2 \rangle \right).
\end{multline}
Comparing with Joule's law, we find the electrical conductivity of the network:
\begin{multline}\label{eq:G-eff}
G^\text{eff}_\text{MFA} = g_\text{m} n \\ \times
\left( \langle l^2 \rangle \langle \cos^2\alpha \rangle + \frac{2}{E} \langle l \delta u \cos\alpha \rangle + \frac{1}{E^2} \langle(\delta u)^2 \rangle \right).
\end{multline}
Here the second and third terms describe corrections to the ``straightforward'' MFA, which does not take potential fluctuations into account.

\subsubsection{Relations for a network corresponding to a 2D Poisson--Voronoi diagram on a plane}

Consider an important special case where the graph corresponds to the edges of a 2D PVD. In this case, the graph is 3-regular. The orientations of its edges are equiprobable, so $\langle\cos^2 \alpha \rangle = 1/2$.

Taking into account the information presented in Section~\ref{subsec:VT}, and since $n = 3 n_\text{s}$,
\begin{equation}\label{eq:PVTmeanl}
 \langle l \rangle = \frac{2}{\sqrt{3 n}},
\end{equation}
\begin{equation}\label{eq:PVTmeanl2}
 \langle l^2 \rangle = \frac{3 C}{n}.
\end{equation}
Then the expression for the electrical conductivity of the original network~\eqref{eq:G-orig} takes the form
\begin{equation}\label{eq:PVTsigmaMFA}
  G^\text{orig} = \lambda \frac{\sqrt{n}}{\sqrt{3}}\left( 1 + \epsilon^\text{orig} \right) \approx 0.57735 \lambda \sqrt{n},
\end{equation}
where
\begin{equation}\label{eq:epsorig}
\epsilon^\text{orig} = \frac{2\sqrt{3n}}{E}\left\langle\delta u \cos\alpha\right\rangle + \frac{\sqrt{3n}}{E^2}\left\langle\frac{(\delta u)^2}{ l }\right\rangle.
\end{equation}
The leading term of Eq.~\eqref{eq:PVTsigmaMFA} coincides with~\eqref{eq:SigmaAPL}, which was obtained in~\cite{Tarasevich2023APL}.

The expression for the electrical conductivity of the effective network is
\begin{equation}\label{eq:PVTsigmaEMT}
  G^\text{eff} = \frac{3 C}{2} g_\text{m}\left( 1 + \epsilon^\text{eff} \right) \approx 0.945 g_\text{m},
\end{equation}
where
\begin{equation}\label{eq:epseff}
\epsilon^\text{eff} = \frac{4 n}{3 C E}\left\langle  l  \delta u \cos\alpha \right\rangle + \frac{2 n}{3 C E^2}\left\langle(\delta u )^2\right\rangle.
\end{equation}

\subsubsection{Estimation of electrical conductivity within the effective-medium theory}

Since the electrical conductance of a linear conductor of constant cross-section is related to its length as $g = \lambda/l$, the PDF of conductance can be expressed in terms of the PDF of edge lengths. Switching to the edge length distribution \(f_L(l,n_\text{s})\) and introducing the dimensionless variable \(\xi = l\sqrt{n_\text{s}}\), \(f_L(l,n_\text{s})\mathrm{d}l = f_L(\xi ;1)\mathrm{d}\xi\), and also \(g = \lambda / l = \lambda \sqrt{n_\text{s}} / \xi\), from the basic EMT equation~\eqref{eq:EMT} for a 2D PVD we obtain:
\begin{equation}\label{eq:IEEMT}
\int\limits_0^\infty \frac{\gamma \xi - 1}{\gamma \xi + 2} f_L(\xi ;1) \, \mathrm{d}\xi = 0,
\end{equation}
where
\begin{equation}\label{eq:gamma}
\gamma = \frac{g_\text{m}}{\lambda \sqrt{n_\text{s}}}
\end{equation}
and where we have used $z=3$.
The integral equation~\eqref{eq:IEEMT} can be solved numerically using the tabulated values of $f_L(l,1)$ given in~\cite{Brakke2005}. The numerical solution yields $\gamma \approx 1.7556$. Thus,
\begin{equation}\label{eq:gmEMT}
g_\text{m} = \gamma \lambda \frac{\sqrt{n}}{\sqrt{3}}.
\end{equation}
Substituting $g_\text{m}$ from Eq.~\eqref{eq:gmEMT} into Eq.~\eqref{eq:PVTsigmaEMT}, we obtain for the leading term
\begin{equation}\label{eq:PVTsigmaEMT1}
  G^\text{eff} = \gamma C \frac{\sqrt{3}}{2} \lambda\sqrt{n} \approx 0.958 \lambda\sqrt{n} .
\end{equation}

\subsection{Results of calculations}

\begin{figure}[!htbp]
  \centering
  \includegraphics[width=\columnwidth]{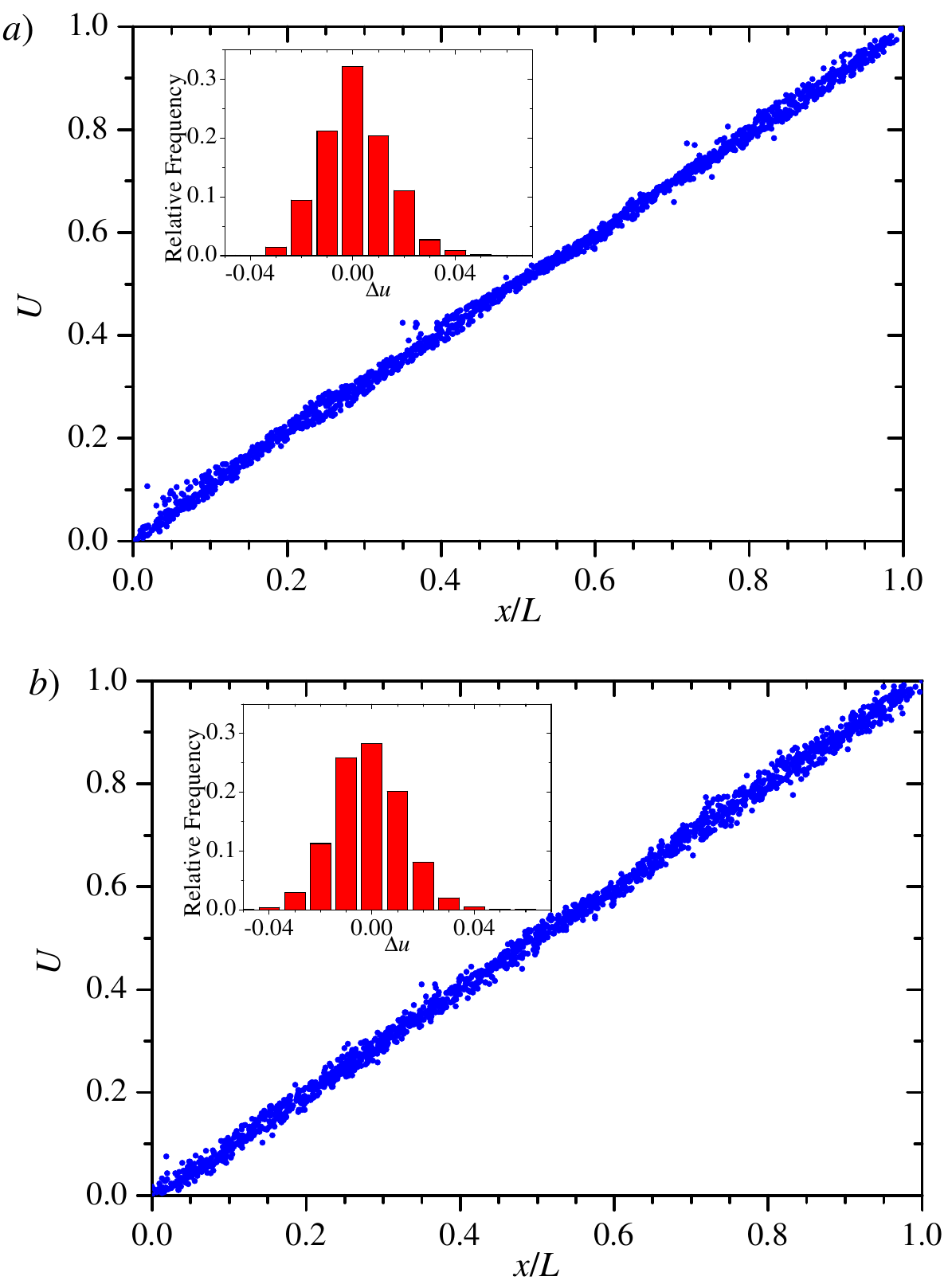}
  \caption{Node potentials as a function of the coordinate in networks corresponding to the edges of a 2D PVD with 1000 seeds. a) Original network, $\lambda = 1$. b) Effective network, $g_\text{m} =1$. Insets: distributions of node potential fluctuations.}\label{fig:Potential}
\end{figure}

Since the MFA is based on the assumption of a linear dependence of the potential on the coordinate, we tested this assumption. Figure~\ref{fig:Potential} presents calculation results confirming the validity of the assumption of a linear potential variation and small fluctuations at the nodes for both the original and effective networks. The fluctuation is understood as the deviation of the potential at a network node with abscissa $x$ from the linear potential $u(x) - U_0 x$.

Additionally, we tested the assumption used in deriving the formulas that edge orientations and lengths in a 2D PVD are independent. Table~\ref{tab:comp} demonstrates that the deviations of the theoretical values from the direct calculation results do not exceed one percent, confirming the appropriateness of our assumption.
\begin{table}[!htbp]
\caption{Comparison of theoretical estimates and direct calculation results intended to verify the independence of edge orientations and lengths in a 2D PVD.}\label{tab:comp}
\begin{ruledtabular}
\begin{tabular}{lll}
network & calculation & theory \\
\hline
original & $\left\langle l \cos^2\alpha \right\rangle = 0.3387$ & $\langle l \rangle \left\langle \cos^2\alpha \right\rangle = 0.3373$ \\
effective & $\left\langle l^2 \cos^2\alpha \right\rangle = 0.3256$ & $\left\langle l^2 \right\rangle \left\langle \cos^2\alpha \right\rangle = 0.3226$ \\
\end{tabular}
\end{ruledtabular}
\end{table}

To verify the correctness of the effective conductivity value~\eqref{eq:gmEMT}, as well as to assess the influence of the width of the distribution~\eqref{eq:appfG} on the accuracy of the MFA, we calculated the electrical conductivity of honeycomb networks in which the edge resistances correspond to the PDF of edge lengths in a 2D PVD~\cite{Brakke2005}. A least-squares fit (LSF) gives
\begin{equation}\label{eq:G-HEX-LSF}
G^\text{hex}\text{LSF} = (0.5686 \pm 0.0023)\lambda\sqrt{n},
\end{equation}
where the quoted uncertainty is the standard error of the fit parameter. This value is approximately 3\% lower than the value predicted by EMT
\begin{equation}\label{eq:G-HEX-EMT}
G^\text{hex}\text{EMT} \approx 0.5852 \lambda \sqrt{n},
\end{equation}
obtained by substituting~\eqref{eq:gmEMT} into~\eqref{eq:Ghex}.

The slight deviation of the EMT prediction from the direct calculation is not surprising, since~\citet{Marchant1979} explained the nature of the error inherent in applying EMT to regular networks. Namely, replacing the distribution of edge conductances with a single value found for the ``effective network'' leads to an error that grows with the broadening of the actual conductance distribution; in other words, the wider the distribution, the larger the error. Figure~\ref{fig:fvsg} shows that the PDF~\eqref{eq:appfG} is broad; therefore, one would expect that the inaccuracy of the EMT prediction would be significant~\cite{Marchant1979}. Thus, the calculation, on the one hand, confirmed that the effective conductance value obtained is correct, and on the other hand showed that the expected error of the EMT prediction for the effective network is a few percent.

\begin{figure}[!htb]
 \centering
 \includegraphics[width=\columnwidth]{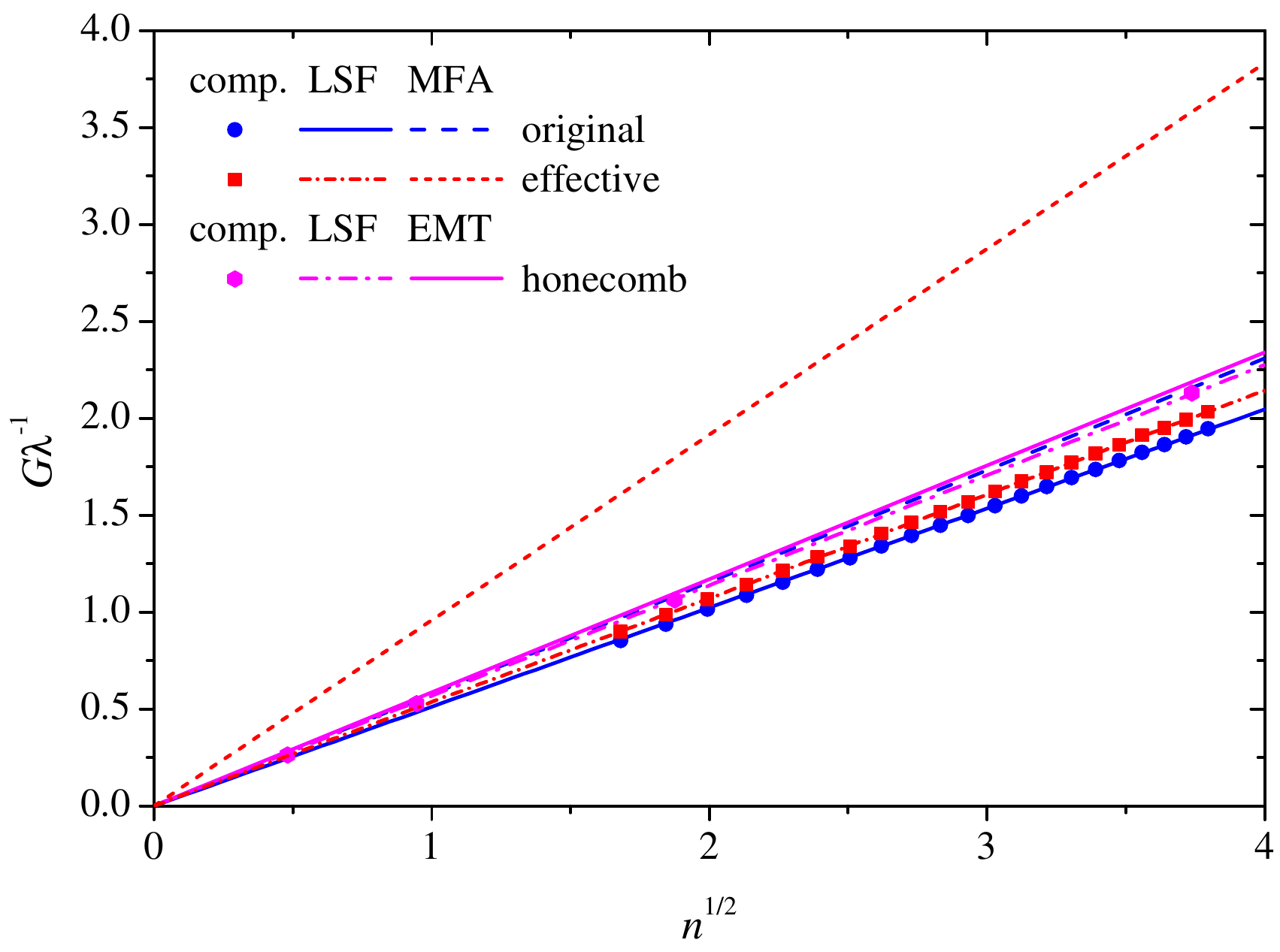}
 \caption{Comparison of direct calculation results for the electrical conductivity of various networks with EMT predictions: for the original and effective networks obtained from a 2D PVD, and for a honeycomb network whose edge conductances obey the distribution $f_G(g)$ shown in Fig.~\ref{fig:fvsg}.}\label{fig:Conductivity}
\end{figure}

We calculated the electrical conductivities of networks obtained from a 2D PVD with a conductance per unit edge length of $\lambda =1$. The results averaged over 50 independent realizations are shown in Fig.~\ref{fig:Conductivity}; the standard error of the mean does not exceed the marker size. An LSF (solid line in Fig.~\ref{fig:Conductivity}) gives
\begin{equation}\label{eq:G-PVT-LSF}
G^\text{orig}_\text{LSF} = (0.5115 \pm 0.0003)\lambda\sqrt{n},
\end{equation}
which is consistent with both the calculation results presented in~\cite{Tarasevich2023APL},
\begin{equation}\label{eq:G-PVT}
 G^\text{PVT} = (0.5087 \pm 0.0027) \lambda \sqrt{n},
\end{equation}
and the MFA estimates~\eqref{eq:PVTsigmaMFA} (dashed line in Fig.~\ref{fig:Conductivity}).

Figure~\ref{fig:Conductivity} also shows the direct calculation results for the electrical conductivity of the effective networks. An LSF (dash-dotted line in Fig.~\ref{fig:Conductivity}) gives
\begin{equation}\label{eq:G-EMT-LSF}
G^\text{eff}_\text{LSF} = (0.5357 \pm 0.0002)\lambda\sqrt{n}.
\end{equation}
However, the MFA estimate~\eqref{eq:PVTsigmaEMT1} (dashed line) is strongly overestimated.

Calculation of the electrical conductivity of effective networks based on a 2D PVD with $g_\text{m}=1$ and various intensities of the Poisson process showed that, in agreement with Eq.~\eqref{eq:PVTsigmaEMT}, the network conductivity does not depend on the edge concentration. On the other hand, the calculated conductivity value is $0.528 \pm 0.001$, which is significantly lower than the predicted value of $0.945$ from~\eqref{eq:PVTsigmaEMT}.

We could not find a theoretical way to estimate the corrections associated with potential fluctuations at the nodes; however, we calculated these corrections numerically. They turned out to be small compared with unity and negative. Specifically, for the original network, including the correction reduced the theoretical estimate~\eqref{eq:PVTsigmaMFA} by 1\% to $G^\text{orig} = (0.57152 \pm 0.00003)\lambda\sqrt{n}$, and for the effective network, the theoretical conductivity estimate~\eqref{eq:PVTsigmaEMT1} including the correction decreased by 3.5\% to $G^\text{eff} = (0.924 \pm 0.001)\lambda\sqrt{n}$. Thus, the corrections reduce the conductivity and, consequently, the thermal power. This is fully consistent with J. C. Maxwell's theorem. It states that in a system of conductors in the absence of internal electromotive forces, the heat generated by currents distributed in accordance with Ohm's law is less than for any other distribution of currents satisfying the actual conditions of current injection and extraction~\cite{Maxwell2016}. On the other hand, these corrections are so small that they cannot explain why the MFA so strongly overestimates the conductivity in the case of the effective network.

To estimate the error introduced into the mean-field approximation when the correction is neglected, we calculated the contributions associated with potential fluctuations at the edge endpoints (Table~\ref{tab:fluctuations}). The results are averaged over 50 independent realizations. Table~\ref{tab:fluctuations} shows that corrections to the standard MFA are not small in the case of the effective network, which leads to the aforementioned significant discrepancy between theoretical estimates of electrical conductivity and the results of direct calculations.
\begin{table}[!htbp]
\caption{Estimates of the correction terms for formulas~\eqref{eq:PVTsigmaMFA} and~\eqref{eq:epseff}, obtained by direct calculation of the potentials at the edge ends. For clarity, the contributions of the first ($\epsilon_1$) and second ($\epsilon_2$) terms are presented.}\label{tab:fluctuations}.
\begin{ruledtabular}
\begin{tabular}{llll}
network & $\epsilon_1$ & $\epsilon_2$ & $\epsilon$ \\
\hline
%
\multicolumn{4}{c}{1000 seeds, $n \approx 2.82$}\\	
original  & -0.241 & 0.121 & -0.120\\
effective & -0.891 & 0.445 & -0.445\\
\multicolumn{4}{c}{2000 seeds, $n \approx 5.71$}\\	
original  & -0.235 & 0.117 & -0.117\\
effective & -0.893 & 0.446 & -0.446\\
\multicolumn{4}{c}{3000 seeds, $n \approx 8.60$}\\	
original  & -0.229 & 0.115 &-0.115 \\
effective & -0.884 & 0.442 & -0.442\\
\multicolumn{4}{c}{4000 seeds, $n \approx 11.5$}\\	
original  & -0.227 & 0.113 & -0.113\\
effective & -0.884 & 0.442 & -0.442\\
\multicolumn{4}{c}{5000 seeds, $n \approx 14.4$}\\	
original  & -0.225 & 0.112 & -0.112\\
effective & -0.885 & 0.442 & -0.442\\
\end{tabular}
\end{ruledtabular}
\end{table}

\section{Conclusion\label{sec:concl}}

Following~\cite{Zeng2020,Kim2022,Tarasevich2023PRE,Esteki2023,Tarasevich2023APL}, we modeled CTB TCFs using random resistor networks whose structure corresponded to the edges of a 2D PVD. We considered two limiting cases: the original network, in which the edge resistance is directly proportional to its length, and the effective network with identical edge conductances obtained using EMT~\cite{Marchant1979}. To assess the dependence of the electrical conductivity of such CTB TCFs on the basic physical parameters, we used the MFA. Comparison of the MFA estimate with direct calculation results shows that the MFA overestimates the conductivity by about 13\% for the original network; for the effective network, the overestimation is about 79\%. We attribute this to the MFA working well for networks in which the conductance of the edges depends only on their length. In this case, the assumption that the currents in individual conductors are independent holds with a reasonable degree of accuracy. For the effective network, however, this assumption becomes completely invalid, leading to a strong overestimation of conductivity within the MFA. The significant overestimation of conductivity within the MFA for the effective network is associated with the structure of the leading term in the Joule heating. Without fluctuations, the current in an edge is proportional to its length. This contradicts the actual current distribution, in which long edges with the same resistance cannot carry a greatly increased current. In real networks, conductors can have different cross-sections and be made of different materials, so the edge resistance does not necessarily depend only on its length, as in the original network considered here. Accordingly, caution should be exercised when applying the MFA to real networks. Note that the two model networks considered here are extreme cases; real CTB TCFs are likely to lie somewhere between them.

In crack templates, different cracks can generally have different widths, although such systems are not directly investigated in the present work. This is particularly relevant for hierarchical crack networks. For such systems, the electrical resistance of individual conductors cannot be assumed to be simply proportional to the conductor length. Therefore, the present results suggest, as a possible implication or future extension of the present work rather than a demonstrated conclusion, that MFA-based estimates of electrical conductivity may be strongly overestimated for CTB TCFs with variable-width cracks. A dedicated study of hierarchical crack networks is required to verify this conjecture.

The result for the honeycomb network with random edge conductances is noteworthy. Although Foster's theorem relies essentially only on all vertex degrees being equal, the EMT estimate for the honeycomb network proved more accurate than for the regular graph corresponding to the edges of a 2D PVD. We attribute this to the fact that the 2D PVD is less homogeneous than the honeycomb graph.


\section*{Acknowledgments}
Yu.Yu.T. thanks Avik Chatterjee for numerous discussions of Foster's theorem and EMT At the very early stage of this work.

\section*{Funding}
A.V.E. acknowledges funding from the Russian Science Foundation, grant No.~25-21-00460.  At the very early stage of this work, during his stay at Instituto de F\'{\i}sica, Universidade Federal Fluminense, Niter\'{o}i, RJ, Brazil, Yu.Yu.T. acknowledges partial financial support from FAPERJ, grants No.~E-26/202.666/2023 and No.~E-26/210.303/2023.

\section*{Roles}
Yuri Y Tarasevich: Supervision (lead), Validation (lead), Visualization (lead), Writing --- original draft (lead), Writing review \& editing (lead), Andrei V Eserkepov: Funding acquisition (lead), Software (lead), Irina V Vodolazskaya: Validation (supporting), Writing --- review \& editing (supporting), Validation (supporting), Writing --- original draft (supporting), Writing --- review \& editing (supporting.

\section*{Data}
The data cannot be made publicly available upon publication because they are not available in a format
that is sufficiently accessible or reusable by other researchers. The data that support the findings of this
study are available upon reasonable request from the authors.

\bibliography{long}

\end{document}